\documentclass[elsart]{article}
\input epsf.sty

\def\lsim{\lower0.6ex\vbox{\hbox{$ \buildrel{\textstyle 
<}\over{\sim}\ $}}}
\def\rsim{\lower0.6ex\vbox{\hbox{$ \buildrel{\textstyle 
>}\over{\sim}\ $}}}
\def\hii{H\thinspace{$\scriptstyle{\rm II}$}~}
\def\hi{H\thinspace{$\scriptstyle{\rm I}$}~}
\def\he3{$^{3}$He}
\def\ie{{\it i.e.},~}
\def\eg{{\it e.g.},~}
\def\nutau{$\nu_\tau$~}
\def\3he{$^3$He}
\def\4he{$^4$He}
\def\6li{$^6$Li}
\def\7li{$^7$Li}
\def\he3{$^3$He}
\def\Yp{Y$_{\rm P}$~}
\def\etal{{\it et al.~}}
\def\hii{H\thinspace{$\scriptstyle{\rm II}$}~}
\def\popii{Pop\thinspace{$\scriptstyle{\rm II}$}~}

\def\ga{\mathrel{\mathpalette\fun >}}
\def\fun#1#2{\lower3.6pt\vbox{\baselineskip0pt\lineskip.9pt
  \ialign{$\mathsurround=0pt#1\hfil##\hfil$\crcr#2\crcr\sim\crcr}}}
\def\beq{\begin{equation}}
\def\eeq{\end{equation}}

\begin{document}
\begin{flushright}
UMN--TH--1802/99\\
TPI--MINN--99/29 \\
astro-ph/9905320 \\
May 1999\\
\end{flushright}
\vskip 0.7in
 
\begin{center} 

{\Large{\bf Primordial Nucleosynthesis:\\
 Theory and Observations}}
 
\vskip 0.4in
{Keith A. Olive$^1$, Gary Steigman$^2$, and Terry P. Walker$^2$}
 
\vskip 0.2in
{\it $^1${Theoretical Physics Institute, School of Physics \& Astronomy,\\
University of Minnesota, Minneapolis, MN 55455}}\\
\vskip 0.1in

{\it $^2${Departments of Physics and Astronomy,
The Ohio State University, \\ 
Columbus, OH 43210, USA}}\\

\vskip 1.0in
{\bf Abstract}
\end{center}

We review the Cosmology and Physics underlying Primordial Nucleosynthesis
and survey current observational data in order to compare the predictions 
of Big Bang Nucleosynthesis with the inferred primordial abundances.  From 
this comparison we report on the status of the consistency of the standard 
hot big bang model, we constrain the universal density of baryons (nucleons),
and we set limits to the numbers and/or effective interactions of hypothetical
new ``light" particles (equivalent massless neutrinos).

\newpage
\noindent
\section{Introduction}

At present, the Universe is observed to be expanding \cite{Hubble} and 
filled with radiation \cite{Penzias} - \cite{cobet} which is very cold
today  (T$_{0} = 2.728$~K \cite{Fixsen}).  If the evolution of such a
Universe is  traced back in time to earlier epochs which were hotter and
denser, the  early Universe is a Primordial Nuclear Reactor during its
first 20 minutes  ($\approx $1000~sec).  As the early Universe expands 
and cools, nuclear  reactions are prematurely aborted before the heavier
elements can be  synthesized.  Only the light nuclides deuterium (D),
helium-3 (\he3),  helium-4 ($^4$He), and lithium-7 ($^7$Li) can be
synthesized in abundances  comparable to those observed (or, observable!)
in a variety of astrophysical  sites (\eg stars; cool, neutral gas; hot,
ionized gas).  Since the relative abundances of the primordially-produced
nuclides depend on the density of  nucleons (baryons) and on the
early-Universe expansion rate, a comparison  of the predicted and observed
abundances provides a key test of the standard  model of cosmology,  as 
well as an indirect
``measurement" of the baryon density of the Universe  which is equally
sensitive to dark and luminous baryons (\ie Is the  early-Universe nucleon
abundance consistent with that inferred today from  non-BBN data?), and
offers a unique probe of hypothetical new particles  (beyond the standard
model) whose presence would have altered the expansion  rate of the early
Universe (hence changing the time available for element  synthesis).  As
one of the pillars of the standard model of Cosmology, BBN  opens a unique
window on the Universe.

In this review, dedicated to the memory of our friend and colleague 
Dave Schramm, we review the basic physics and cosmology relevant to 
the calculation of the primordial yields, both in the standard model 
and in simple extensions of the standard model.  Then we compare the 
current predictions, based on up-to-date nuclear and weak interaction
rates, with the primordial abundances inferred from observational data
obtained from a variety of astrophysical sites using a variety of
astronomical techniques.  Since the BBN-prediction part is relatively
simple and straightforward, it is the data which lies at the core of 
such comparisons.  The good news is that the key nuclides are observed 
in a variety of objects using very different techniques, thus minimizing 
correlated systematic errors in the abundance determinations.  Additional 
good news is that modern telescopes and detectors have provided high 
quality data whose statistical errors have been shrinking dramatically.  
The bad news is that, for most abundance determinations, the accuracy 
is now limited by our ignorance of possible systematic errors which are 
often difficult to quantify using extant data alone.  Therefore, a large 
part of this review is devoted to the data and our assessment of the 
uncertainties.  In this, we strive to err on the side of caution.  When 
mutually contradictory data appears (as it does for primordial deuterium) 
we will explore the consequences of each option, letting the reader draw 
his/her own conclusion.  Given the rapid pace of observational cosmology 
at present, the quantitative abundances derived from current data are 
likely ephemeral.  However, it is our hope that our discussion here will 
set the stage for any changes new data will provide.

\section{Primordial Nucleosynthesis}


All that is needed to predict the primordial abundances of the light elements 
within the context of the standard models of cosmology and particle physics 
is the current temperature and expansion rate of the Universe.  Then, under 
the assumptions that the Universe is homogeneous and isotropic and that the 
standard model of particle physics is the correct description of the particle 
content of the Universe at temperatures of order a few MeV, the predicted 
primordial abundances of D, \3he, \4he, and \7li depend only on the baryon 
density. That is, the predictions of standard BBN are uniquely determined 
by one parameter, $\eta$, the baryon-to-photon ratio: $\eta_{10} = 273\Omega_
{\rm B}h^{2}$ ($\Omega_{\rm B}$ is the ratio of the baryon density to the 
critical density and the Hubble parameter is H$_{0}$ = 100h km/s/Mpc; $\eta_
{10} = 10^{10}\eta$).  

The primordial yields of light elements are determined by a competition 
between the expansion rate of the Universe, the rates of the weak interactions 
that interconvert neutrons and protons, and the rates of the nuclear reactions 
that build up the complex nuclei.  Neglecting the contributions of curvature 
and the cosmological constant, which are small in the early Universe, the 
expansion rate is determined by the Friedmann equation: 
\beq
	H^2  \equiv \left({\dot{R} \over R}\right)^2  \approx 
\frac{8 \pi}{3} G_{N} \rho
\label{H}
\eeq
where $R$ is the scale factor.
For standard BBN the energy density, $\rho$, at the time nucleosynthesis 
begins (about 1 second after the Big Bang) is described by the standard 
model of particle physics 
\beq 
\rho =\rho_\gamma + \rho_e + {\rm N}_\nu \rho_\nu
\eeq
where $\rho_\gamma$, $\rho_e$, and $\rho_\nu$ are the energy density of 
photons, electrons and positrons, and massless neutrinos and anti-neutrinos 
(one species), respectively, and $N_\nu$ is the equivalent number of 
massless neutrino species which, in standard BBN, is exactly 3.

At high temperatures, neutrons and protons can interconvert via weak 
interactions:
$n + e^+  \leftrightarrow   p + {\bar \nu_e},~
n + \nu_e  \leftrightarrow   p + e^- ,$ and $
n   \leftrightarrow   p + e^- + {\bar \nu_e}.$ 
As long 
as the interconversion rate of neutrons and protons is faster than the 
expansion rate, the neutron-to-proton ratio tracks its equilibrium value, 
exponentially decreasing with temperature.  This condition holds for 
temperatures $T \ga 1$ MeV as can be seen from a comparison of estimates 
of the weak rates
 $\Gamma_{\rm wk} =  n \langle \sigma v \rangle { \sim  }~0(10^{ -  2}) T^{
5}/M_{ W}^4,
$
 and the expansion rate
$ H  =  \left({8 \pi {}G_{ N}  \rho {}/ 3}\right)^{ 1/2}       
 \sim 5.4\  T^{ 2}/M_{ P},$ where $M_{ W}$  and $M_{ P}$ are the electroweak
 and Planck masses, respectively\footnote{A detailed numerical calculation 
would find that equilibrium is maintained down to 0.8 MeV.}. 
Once the interconversion rate becomes less than the expansion rate, n/p 
effectively ``freezes-out'' (at about 1/6), thereafter decreasing slowly 
due to free neutron decay. 

Although freeze-out occurs at a temperature below the deuterium
binding energy, $E_B = 2.2$ MeV, the first link in the nucleosynthetic 
chain, $p+n \rightarrow$ D $+~\gamma$, is not effective since the 
photo-destruction rate of deuterium ($\propto n_\gamma e^{-E_B/T}$) 
is much larger than the production rate ($\propto n_B$) due to the
large  photon-to baryon ratio ($\ga 10^9$).  As soon as deuterium 
becomes stable against photodissociation ($\sim 80$ keV) neutrons 
are bound up into \4he with an efficiency of 99.99\%, driven by 
the stability of the \4he nucleus. By this time, n/p has dropped to 
$\sim 1/7$, and simple counting yields an estimated \4he mass fraction
\beq
Y_{\rm P} \approx {2(n/p) \over \left[ 1 + (n/p) \right]} = 0.25.
\label{ynp}
\eeq 
In addition, the large binding energy of \4he insures that its primordial 
abundance is relatively insensitive to the nuclear reaction rates (which 
increase with increasing baryon density ($\eta$)).  D (and \3he) is burned 
to get to  complex nuclei and thus its abundance decreases rapidly with 
increasing $\eta$, making D the perfect baryometer (see section 3.1).  At 
low $\eta$, \7li is destroyed by protons with an efficiency that increases 
with $\eta$, while at relatively high $\eta$, $^7$Be (the dominant route 
to \7li  by subsequent electron capture) is produced more efficiently with 
increasing $\eta$.  Hence the ``Li valley'' in a \7li vs. $\eta$ plot.  
Increasing Coulomb barriers and a lack of stable nuclei at $A = 5$ and 8 
cause standard BBN to struggle to produce \7li and be even less effective 
beyond that.

In Figures 1 -- 3,  the primordial abundances predicted by standard BBN 
are shown as a function of $\eta$.  The width of each curve reflects the 
2$\sigma$ uncertainty in the predictions that results from a Monte Carlo 
analysis of uncertainties in the neutron lifetime and nuclear reaction 
rates \cite{hata}.  The neutron lifetime was taken to be $\tau_n = 887 \pm  
2$ seconds\footnote{The current world average is  $\tau_n = 886.7 \pm  1.9$ 
\cite{rpp} leading to predictions indistinguishable from those displayed.}.   
At $\eta = 5 \times 10^{-10}$ the fractional uncertainties due to $2\sigma$  experimental errors are 0.4\% for Y, 15\% for D or \3he, and 42\% for \7li 
if the errors in the cross sections from Smith \etal \cite{skm} are used.   
We note that recent work \cite{betal}, based on a preliminary reanalysis 
and update of the relevant reaction cross sections, claims smaller 
uncertainties in D and \7li by roughly a factor of two.  The robustness 
of the BBN predictions is directly related to the fact that, for the most 
part, the astrophysical S-factors are measured at energies relevant to 
the BBN environment. 

\begin{figure}[ht]
	\centering
	\epsfysize=3.2truein 
\epsfbox{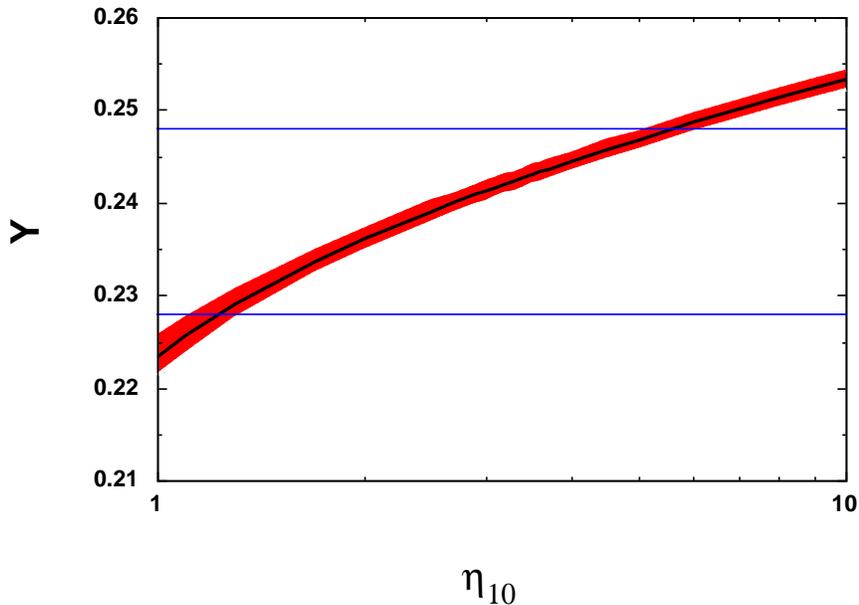}
	\caption{The predicted \4he abundance (solid curve) and the $2 \sigma$
theoretical uncertainty \protect\cite{hata}. The horizontal lines show the
range indicated by the observational data.}
	\label{fig1}
\end{figure}

\begin{figure}[ht]
	\centering
	\epsfysize=3.2truein 
\epsfbox{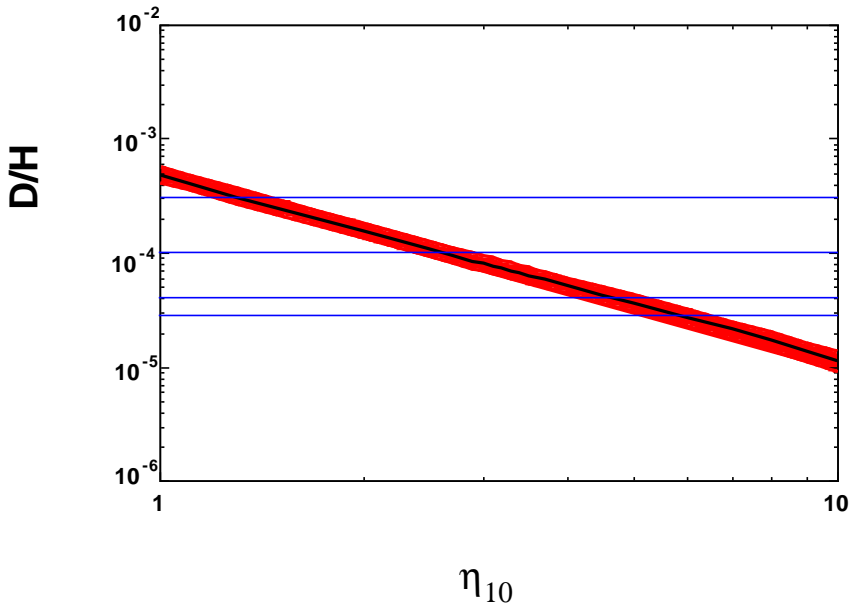}
	\caption{The predicted D/H abundance (solid curve) and the $2 \sigma$
theoretical uncertainty \protect\cite{hata}. The horizontal lines show the
range indicated by the observational data for both the high D/H (upper two
lines ) and low D/H (lower two lines).}
	\label{fig2}
\end{figure}

\begin{figure}[ht]
	\centering
	\epsfysize=3.2truein 
\epsfbox{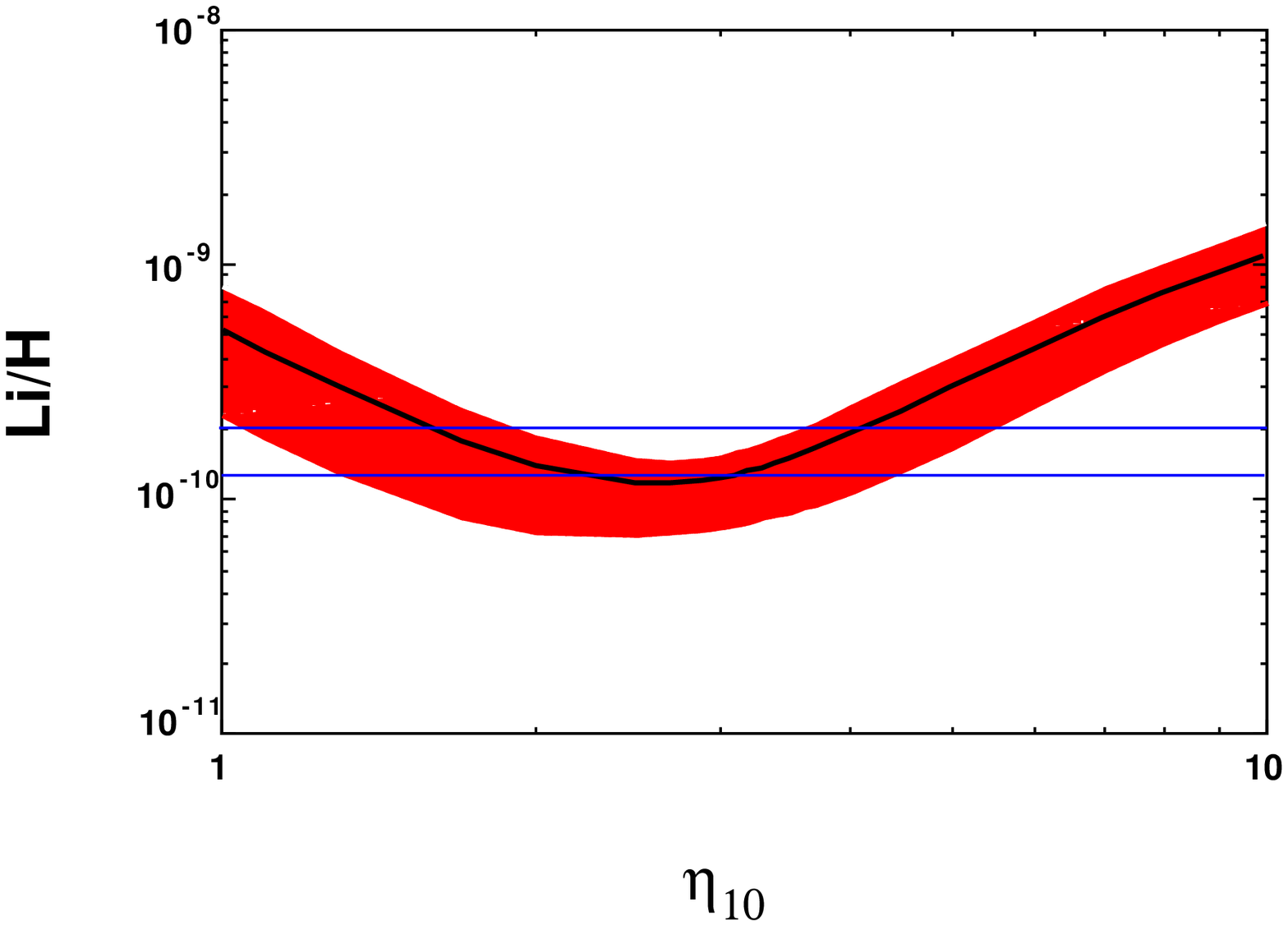}
	\caption{The predicted \7li abundance (solid curve) and the $2 \sigma$
theoretical uncertainty \protect\cite{hata}. The horizontal lines show the
range indicated by the observational data.}
	\label{fig3}
\end{figure}

The fractional uncertainty in the predicted mass fraction of \4he due to 
experimental errors in the reaction rates is, for the range of nucleon 
density of interest, almost entirely due to the uncertainty in the neutron 
lifetime (which translates into an overall uncertainty in the weak 
interconversion rates).  In the last few years considerable effort has 
gone into understanding the theoretical uncertainty in the predicted 
abundance of \4he due to the treatment of the weak interaction rates.  
The BBN code used for the results presented in Walker \etal (WSSOK) 
\cite{wssok} was basically the `Wagoner Code' \cite{wag} (updated first by 
Yang \cite{yssr,ossty,ytsso}) with modifications by Walker\cite{tpw} to include zero 
and finite-temperature radiative corrections and Coulomb corrections to the 
weak rates (as described in Dicus \etal \cite{dicus}). Subsequent modifications included 
an update and enlargement of the nuclear network by Thomas in 
TSOF \cite{tsof}.  Kernan re-examined the code\cite{pjk}, updating the TSOF 
code to include finite nucleon mass effects (as described in Seckel \cite{seck} 
and Lopez, Turner, and Gyuk \cite{ltg}) and finding a relatively large time-step error.   
He also estimated the uncertainty in the \4he mass fraction due to choice of 
finite-temperature prescription and non-equilibrium  effects in the  neutrino 
sector to be $\sim  10^{-4}$.  Kernan's recommendation was, to a level of 
accuracy of a few parts in $10^4$, to simply use the \Yp found in the `WSSOK 
Code' and add 0.003.  This was adopted in Hata \etal \cite{hata} to yield \Yp 
= 0.2467 (theoretical uncertainty of a few parts in 10$^{-4}$) at $\eta=5\times 
10^{-10}$ and $\tau_n = 887$ seconds.  With several versions of the BBN code 
floating around, no one but the owners (and sometimes not even they) knew how 
the various ``\Yp-corrections" were handled and no one had built an independent 
version of the code that contained all these corrections in a self-consistent 
way.  Lopez and Turner \cite{lt} have recently done just that.  Starting
from  scratch, including all the effects mentioned above, and adding
order-$\alpha$  QED corrections (as described by Heckler \cite{heck}) and
detailed non-equilibrium  neutrino effects (as described by Dodelson
and Turner \cite{dt} they find: \Yp =  0.2460 $\pm 0.0002$(theory) at 
$\eta=5\times 10^{-10}$ and $\tau_n  = 885.4$ seconds (if they use $\tau_n 
= 887$ seconds they find \Yp =   0.2467 (Lopez, Private Communication to 
G.S.)).  Indeed, over the entire range $1 \leq \eta_{10} \leq 10$, the 
difference in predicted \4he mass fraction between our code and the Lopez/Turner 
code is 0.0001 $\pm 0.0001$.


\section{From Observations To Primordial Abundances}

To test the standard model it is necessary to confront the predictions 
of BBN with the primordial abundances of the light nuclides which are
not ``observed", but are inferred from observations.  The path from
observational data to primordial abundances is long and twisted and 
often fraught with peril.  In addition to the usual statistical and
insidious systematic uncertainties, it is necessary to forge the
connection from ``here and now" to ``there and then", \ie to relate the
derived abundances to their primordial values.  It is fortunate that
each of the key elements is observed in different astrophysical sites
using very different astronomical techniques and that the corrections
for chemical evolution differ and, even more important, can be minimized.
For example, deuterium is mainly observed in cool, neutral gas (\hi regions) 
via resonant UV absorption from the ground state (Lyman series), while 
radio telescopes allow helium-3 to be studied via the analog of the 21~cm 
line for $^3$He$^{+}$ in regions of hot, ionized gas (\hii regions).  
Helium-4 is probed via emission from its optical recombination lines in 
\hii regions.  In contrast, lithium is observed in the absorption spectra 
of hot, low-mass halo stars.  With such different sites, with the mix 
of absorption/emission, and with the variety of telescopes involved, 
the possibility of correlated errors biasing the comparison with the 
predictions of BBN is unlikely.  This favorable situation extends to the 
obligatory evolutionary corrections.  For example, although until recently 
observations of deuterium were limited to the solar system and the Galaxy, 
mandating uncertain corrections to infer the pregalactic abundance, the 
Keck and Hubble Space telescopes have begun to open the window to deuterium 
in high-redshift, low-metallicity, nearly primordial regions (Lyman-$\alpha$ 
clouds).  Observations of $^4$He in low-metallicity ($\sim 1/50$ of solar) 
extragalactic \hii regions permit the evolutionary correction to be reduced 
to the level of the statistical uncertainties.  The abundances of lithium 
inferred from observations of the very metal-poor halo stars (one-thousandth 
of solar and even lower) require almost no correction for chemical evolution.  
On the other hand, the status of helium-3 is in contrast to that of the other 
light elements. Although all prestellar D is converted to \he3 during pre-main 
sequence evolution, \he3 is burned to $^4$He and beyond in the hotter 
interiors of most stars, while it survives in the cooler exteriors.  For lower
mass stars a greater fraction of the prestellar \he3 is expected to survive 
and, indeed, incomplete burning leads to the buildup of \he3 in the interior 
which may, or may not, survive to be returned to the interstellar medium 
\cite{rst}.  In fact, some planetary nebulae have been observed to be highly 
enriched in \3he, with abundances \3he/H $\sim 10^{-3}$ \cite{rood}.  Although 
such high abundances are expected in the remnants of low mass stars \cite{rst, 
dst}, if all stars in the low mass range produced comparable abundances, we 
would expect solar and present ISM abundances of \3he to greatly exceed their 
observed values \cite{dst,galli,orstv}.  It is therefore necessary that at 
least {\it some} low mass stars are net destroyers of \he3.  For example, there 
could be ``extra" mixing below the convection zone in these stars when they are 
on the red giant branch \cite{char,gal,osst}.  Given such possible complicated histories of survival, destruction, and production, it is difficult to use the 
current Galactic and solar system data to infer (or, even bound) the primordial abundance of \he3.  For this reason, we will not consider \3he any further in 
this review.

The generally favorable observational and evolutionary state of affairs for 
the nuclides produced during BBN is counterbalanced by the likely presence 
of systematic errors which are difficult to quantify and, in some cases, by 
a woefully limited data set.  As a result, although cosmological abundance determinations have taken their place in the current ``precision" era, it is 
far from clear that the present abundance determinations are ``accurate".  
Thus, the usual {\it caveat emptor} applies to any conclusions drawn from 
the comparison between the predictions and the data.  With this caution in 
mind we survey the current status of the data to infer ``reasonable" ranges 
for the primordial abundances of the key light elements.

\subsection{Deuterium}

Deuterium is the ideal baryometer.  As we have noted above the BBN-predicted 
D/H ratio is a strong function of the baryon-to-photon ratio $\eta$.  A
determination of the primordial abundance to 10\%, leads to an $\eta$ 
determination accurate to $\sim 6$\%.  Furthermore, since deuterium 
is burned away whenever it is cycled through stars, and there are no 
astrophysical sites capable of producing deuterium in anywhere near 
its observed abundance \cite{els}, any observed D-abundance provides a 
{\it lower} bound to its primordial abundance. Thus, without having to 
correct for Galactic evolution, the deuterium abundance inferred from UV 
observations of the local interstellar medium (LISM) \cite{lin}, D/H = 
$(1.5 \pm 0.1)\times 10^{-5}$ (unless otherwise noted, observational 
errors are quoted at 1$\sigma$), bounds the primordial abundance from 
below and the baryon-to-photon ratio from above \cite{rafs}.  This value 
represents an average along 12 lines of sight in the LISM.  Although they
are not directly relevant to BBN, it is interesting to note that there have 
been several reports \cite{vm,or} of a dispersion in ISM D/H abundances.  
It is not clear whether such variations are related to those inferred 
for the \3he/H abundances in Galactic \hii regions \cite{bbbrw}.

Solar system observations of \he3 permit an indirect determination of 
the pre-solar system deuterium abundance (Geiss \& Reeves 1972).  This 
estimate of the Galactic  abundance some 4.5~Gyr ago, D/H = $(2.1 \pm 
0.5)\times 10^{-5}$ (Geiss \&  Gloeckler 1998), while having larger 
uncertainty, is consistent with the LISM value. There has also been a 
recent measurement of deuterium in the atmosphere of Jupiter using the 
Galileo Probe Mass Spectrometer \cite{jup}, which finds D/H $= (2.6 \pm 
0.7) \times 10^{-5}$.

To further exploit the solar system and/or LISM deuterium determinations 
to constrain/estimate the primordial abundance would require corrections 
for the Galactic evolution of D.  Although the simplicity of the evolution 
of deuterium (only destroyed) suggests that such correction might be very 
nearly independent of the details of specific chemical evolution models, 
large differences remain between different estimates \cite{scov,chiap}.  
It is therefore fortunate  that data on D/H in high-redshift, low-metallicity
Lyman-$\alpha$ absorbers  has become available in recent years
\cite{quas1}-\cite{hity}.  It is expected that such systems still retain
their original,  primordial deuterium, undiluted by the deuterium-depleted
debris of any  significant stellar evolution.  That's the good news.  The
bad news is  that, at present, there are D-abundance determinations claimed
for only  four such systems and that the abundances inferred for two of
them appear  to be inconsistent with the abundances determined in the other
two.  Here  is a prime example of ``precise"  but possibly inaccurate
cosmological  data.  There is a serious obstacle inherent to using
absorption spectra  to measure the deuterium abundance since the
isotope-shifted deuterium  lines are indistinguishable from
velocity-shifted hydrogen.  Such  ``interlopers" may have been  responsible
for some of the early claims 
\cite{quas1} of a ``high" deuterium abundance \cite{steig}.  Data reduction 
errors may have been the source of another  putative high-D system.  At 
present it seems that only three good candidates  for nearly primordial
deuterium have emerged from ground- and space-based  observations.

The absorption system at $z = 3.572$ towards Q1937-1009 was first studied 
by Tytler, Fan \& Burles \cite{quas2} who derived a low D/H = $(2.3 \pm 
0.3 \pm 0.3) \times 10^{-5}$.  Since an uncertain hydrogen column density, 
due to the saturated Lyman series profiles, was the largest source of 
uncertainty \cite{swc}, new, high quality, low-resolution spectra were 
obtained \cite{bty}, which, along with a new fitting procedure led to 
a revised abundance:  D/H = $(3.3 \pm 0.3) \times 10^{-5}$; notice the 
rather poor overlap with the original abundance.  The $z = 2.504$ 
absorption system towards Q1009+2956 provides another potentially accurate 
D-abundance determination \cite{bt2} D/H = $(4.0 \pm 0.7) \times 10^{-5}$.  
There are two other systems studied by Burles \& Tytler (1998) whose derived
D-abundances are consistent with these two, but whose uncertainties are much 
larger.  The weighted mean of the two accurate D-abundance determinations 
leads to a 95\% confidence range: 2.9 $\times 10^{-5} \leq$~D/H~$\leq 4.0 
\times 10^{-5}$.  We adopt this range in our comparisons with the BBN 
predictions.

We note that Levshakov, Kegel \& Takahara (\cite{lkt}, LKT) have used the 
data in \cite{quas2} for the $z = 3.572$ system towards Q1937-1009, but 
with a different model for the velocity distribution of the absorbing gas, 
to derive a (95\% confidence) range 3.5 $\times 10^{-5} \leq$~D/H~$\leq 5.2 
\times 10^{-5}$, which argues for a slightly higher abundance than suggested 
by the Burles \& Tytler \cite{bty} range.  These same authors also used 
their model to reanalyze the Burles \& Tytler \cite{bt2} data for Q1009+2956 
\cite{lkt2} and they derive a 95\% estimated range of 2.9 $\times 10^{-5} \leq$~D/H~$\leq 4.6 \times 10^{-5}$, now in excellent agreement with the 
Burles \& Tytler \cite{bt2} value for this system.  Recently, Levshakov, 
Tytler \& Burles (\cite{lev}, LBT) have joined forces to apply this 
different model to a reanalysis of the $z = 2.504$  absorption system 
towards Q1009+2956, finding a consistent but slightly higher range (68\%): 
D/H $\simeq (3.5 - 5.0) \times 10^{-5}$.

Although deuterium in the two high-redshift absorbers is consistent with a 
primordial abundance in the range 2.9 $\times 10^{-5} \leq$~D/H~$\leq 4.0 
\times 10^{-5}$ (or slightly higher accounting for the LKT and LBT analyses
of the same data),  the deuterium abundance derived 
for the one low-redshift absorber, the $z = 0.701$ system towards Q1718+4807 
observed with the GHRS on HST is significantly different.  This data was
first analyzed by Webb \etal 
\cite{webb} who derived a very high deuterium abundance: D/H = $(20 \pm 5) 
\times 10^{-5}$.  In contrast, LKT \cite{lkt3} using the same data but 
their model for the velocity distribution of the absorbing gas, derive 
an abundance closer to those for the high-redshift absorbers: 4.1 $\times 
10^{-5} \leq$~D/H~$\leq 4.7 \times 10^{-5}$.  Recently, Tytler \etal 
\cite{hity} use new Keck spectra 
to supplement the data from HST to derive a 95\% range: 8 $\times 10^{-5} 
\leq$~D/H~$\leq 57 \times 10^{-5}$, consistent with the Webb \etal (1997) 
estimate.  Clearly the high-D abundance inferred from some analyses of this 
system are inconsistent with the low-D abundances derived from the other two, 
higher-redshift systems.  The sense of the  discrepancy is puzzling since 
it is expected that the deuterium abundance should only decrease with time 
(decreasing redshift).  If, in fact, the high abundance is representative 
of the primordial value, then the other two absorbers should consist of gas 
most of which has been cycled through stars.  The high redshifts and low 
metallicities of these systems suggest this is unlikely. If high D-abundances 
at high-$z$ and low-$Z$ are common, many systems like Q1718+4807 should 
present themselves for analysis.  Tytler (1998) has argued that the absence 
(so far) of very many possible candidates suggests that either the 
abundance determination in the Q1718+4807 absorber is unreliable, or 
the Q1718+4807 absorber is anomalous.

In anticipation of new data which may resolve this conundrum, we prefer 
to keep our options open and discuss the consequences of either of two 
(mutually exclusive) possibilities.  For the low-D case we use the two 
high-$z$ systems and adopt the Burles-Tytler 95\% range: 2.9 $\times 
10^{-5} \leq$~D/H~$\leq 4.0 \times 10^{-5}$.  For the high-D case we 
adopt the range: 1 $\times 10^{-4} \leq$~ D/H~$\leq 3 \times 10^{-4}$ 
based on the $2\sigma$ range of Webb \etal \cite{webb}.  With account 
for the uncertainty in the BBN-predicted D-abundance at fixed $\eta$, 
the lower bound to primordial D/H for the low-D case leads to an upper 
bound to $\eta$ of: $\eta_{10} \leq 6.3$, while the upper bound on D/H 
leads to a lower bound on $\eta$ of: $\eta_{10} \geq 4.2$.  For the high-D 
option, the  corresponding range in $\eta$ is: 1.2 $\leq \eta_{10} \leq 2.8$.  
In making  these estimates we have been ``overly generous" in the sense 
that the $\eta$  values correspond to the ``2$\sigma$" uncertainties in the 
observational data  and the ``2$\sigma$" uncertainties in the BBN predictions.

\subsection{Helium-4}

As the second most abundant nuclide in the Universe (after hydrogen), the
abundance of $^4$He can be determined to high accuracy at sites throughout
the Universe.  To minimize the uncertainty inherent in any correction for
the debris of stellar evolution, it is sensible to concentrate on the data
from low-metallicity, extragalactic \hii regions \cite{p}-\cite{fdo2}.  
Since each data set contains  of order 40 regions, various analyses achieve
statistical uncertainties in their estimate of the primordial helium mass
fraction $\leq 0.003$ (or, $\leq 1\%$).  Further, since the most metal-poor 
of these regions have metallicities of order 1/50 -- 1/30 of solar,
the extrapolation from  the lowest metallicity regions to truly primordial
introduces an uncertainty  no larger than the statistical error.  Although
$^4$He has already entered the  era of ``precision cosmology", difficult to
constrain systematic uncertainties dominate the error budget.  For example,
using published data for 40 low-metallicity regions (excluding the suspect 
NW region of IZw18), Olive \&  Steigman (OS) \cite{osa} find: Y$_{\rm P} = 
0.234 \pm 0.003$ based on the data in \cite{p,evan}.  In contrast, from an
independent data set of 45 low-metallicity regions with only slight overlap 
with that of OS, Izotov \& Thuan (IT) \cite{iz2} infer Y$_{\rm P}  = 0.244 
\pm 0.002$.  Clearly these results are statistically inconsistent.  Several
contributions to this discrepancy can be identified.  Since the intensity 
of the helium recombination emission lines can be enhanced by collisional
excitation \cite{coll}, corrections for collisional excitation are mandatory. 
In \cite{iz,iz2} an attempt was made to use helium-line data alone (5 lines) 
to make this correction, in contrast to the traditional approach using 
information on the electron density derived from non-helium line data 
(see Skillman, Terlevich \& Terlevich \cite{stt} for a discussion).  It 
is of great value that Izotov \etal \cite{iz} (ITL) and IT also analyze 
their data according to the traditional approach since this permits an 
estimate of the effect of their approach on the inferred primordial abundance.  
Using their data for 44 regions analyzed similarly to the data employed in 
OS, they would have derived Y$_{\rm P} = 0.241 \pm 0.002$, reducing the 
discrepancy between OS and IT.  Another source of systematic difference 
between the two analyses can be identified.  By relying on helium (and hydrogen) 
recombination lines, any neutral helium (or hydrogen) present in the \hii 
regions is invisible and must be corrected for.  Since any such correction
will be model dependent and uncertain, Pagel \etal \cite{p} restricted their
attention to \hii regions of ``high excitation" for which this correction
should be minimized.  As a result they (and most of the data utilized by OS)
make no ionization correction.  In contrast ITL, through a misreading of
published models of \hii regions, make a correction for neutral helium while
ignoring the (predicted) larger correction for neutral hydrogen in regions
ionized  by hot stars (metal-poor stars are hotter than the corresponding 
solar metallicity stars).  Skillman, Terlevich \& Terlevich \cite{stt} 
estimate the size of this correction to be of order 1\% ($\Delta $Y$_{\rm P} 
\approx -0.002$), further reducing the discrepancy between the IT and OS 
Y$_{\rm P}$ estimates to $\approx 0.005$ rather than the original 0.010.  
Although IT eliminate the erroneous ionization correction from ITL in their 
more recent work, they actually derive a {\it higher} helium abundance.
IT remark that this may be due to the higher
temperatures in their new regions (compared to the ITL data set).

At present potentially the most significant systematic uncertainty affecting
the derived primordial abundance of helium appears to be that due to possible
underlying stellar absorption (ITL; IT; Skillman, Terlevich \& Terlevich 
\cite{stt}).  It has become clear that the helium abundance determination 
in the NW region of IZw18 is likely contaminated by such absorption, resulting 
in an underestimate of the true abundance.  Other regions in the OS and 
Olive, Skillman \& Steigman \cite{ost3} (OSS) data sets may suffer similar 
contamination, biasing their estimate of the primordial helium abundance to 
values which may be too low.  In contrast, ITL/IT select their regions on 
the basis of the strength of the helium  lines, avoiding those weak-lined 
regions which may be contaminated by underlying stellar absorption.  If, 
indeed, they have been successful in avoiding this systematic error, their 
higher abundance estimate may be closer to the true value.  But, through 
such selection they have run the risk of introducing a bias against finding 
low helium abundances.  

It is clearly crucial that high priority be assigned to using the \hii 
region observations themselves to estimate/avoid the systematic errors 
due to underlying stellar absorption, to collisional excitation and, to 
corrections for neutral helium and/or hydrogen.  Until then, the error 
budget for Y$_{\rm P}$ is likely dominated by systematic rather than 
statistical uncertainties and it is difficult to decide between OS (and 
OSS) and IT.  When account is taken of systematic uncertainties, they may
in fact be consistent with each other.  Therefore, in what follows, we will 
will adopt a generous ``95\%" range of 0.228 $\leq$ Y$_{\rm P} \leq 0.248$
({\it cf,} \cite{fdo2}).

\subsection{Lithium-7}

Cosmologically interesting lithium is observed in the \popii halo 
stars \cite{sp}-\cite{rnb} which are so metal-poor they provide a 
sample of more nearly primordial material than anything observed 
anywhere else in the Universe; the most metal-poor stars have less 
than one-thousandth the solar metallicity.  Of course these halo 
stars are the oldest stars in the Galaxy and, as such, have had 
the most time to modify their surface abundances.  So, although 
any correction for evolution modifying the lithium abundance may be 
smaller than the statistical uncertainties of a given measurement, 
the systematic uncertainty associated with  the dilution and/or 
destruction of surface lithium in these very old  stars could dominate 
the error budget. Additional errors are associated with the modeling 
of the surface layers of these cool, low-metallicity, low-mass stars, 
such as those connected with stellar atmosphere models and the temperature 
scale. It is also possible that some of the observed Li is non-primordial, 
(\eg that some of the observed Li may have been produced by spallation 
or fusion in cosmic-ray collisions with gas in the ISM \cite{sw, wssof}). 

There now exists a very large data set of lithium abundances measured 
in the warmer ($T > 5800K$), metal-poor ([Fe/H] $< -1.3$) halo stars.
Within the errors, these abundances define a plateau (the ``Spite-plateau")
in the lithium abundance -- metallicity plane.  Depending on the choice 
of stellar-temperature scale and model atmosphere the abundance level 
of the plateau is: A(Li) $\equiv 12 + $log(Li/H)$ = 2.2 \pm 0.1$, with 
very little intrinsic dispersion around this plateau value (\eg \cite{mol}).  
This small dispersion provides an important constraint on models which 
attempt to connect the present surface lithium abundances in these stars 
to the original lithium abundance in the gas out of which these stars 
were formed some 10 -- 15 Gyr ago.  ``Standard" (\ie non-rotating) 
stellar models predict almost no lithium depletion and, therefore, 
almost no dispersion about the Spite-plateau \cite{del}.

Early work on mixing in models of rotating stars was very uncertain,
predicting as much as an order of magnitude \7li depletion.  Recently, 
Pinsonneault \etal \cite{pin1}, building on progress in the study of the
angular momentum evolution of low-mass stars \cite{krish}, constructed 
stellar models which reproduce the angular momentum evolution observed 
for low-mass open cluster stars, and have applied these models, normalized 
to the open cluster data and to the observed solar lithium depletion, to 
the study of lithium depletion in main sequence halo stars.  Using the 
distribution of initial angular momenta inferred from young open clusters 
for the halo stars leads to a well-defined lithium plateau with modest 
scatter and a small population of ``outliers" (overdepleted stars) which is 
consistent with  the data.  Consistency with the solar lithium, with the 
open cluster stars, and with the (small) dispersion in the Spite-plateau 
may be achieved for  depletion factors between 0.2~dex and 0.4~dex \cite{pin1}.

The amount of depletion can also be limited \cite{sfosw,fdo4,vf} by 
observations of \6li \cite{li6}.  If the original \6li in halo stars
is assumed to be as high as the solar value, an upper bound of 0.4 dex 
\7li depletion in rotational models is obtained from \6li data \cite{pin1}.  
Recent analysis \cite{fdo4} suggests a more stringent ({\it albeit} model 
dependent) \7li depletion limit of 0.2 dex based on constraints on the 
low metallicity ([Fe/H] $\approx -2.3$) production of \6li.  Clearly \6li 
plays a vital role when it comes to constraining \7li depletion -- the key 
issue to be resolved is the evolution of \6li in low metallicity environments 
and the data required are the simultaneous observations of the isotopes of 
Li, Be, and B in low metallicity halo stars.

  
Very recently, Ryan, Norris \& Beers (RNB) \cite{rnb} have presented data 
for 23 very metal-poor ([Fe/H] $\lsim -2.5$) field turnoff stars, chosen 
specifically to lie in a limited range of metallicity so as to facilitate 
the study of the dispersion in the Spite plateau.  Although the limited 
data set subjects any conclusions to the uncertainties due to small number 
statistics, these data confirm previous suggestions \cite{mol} that there 
is very little dispersion about the plateau abundance.   RNB claim
evidence for a slope in the A(Li) vs [Fe/H]  data (an increase of Li 
with Fe). If real, this suggests that not all of the inferred lithium 
is primordial.  In a recent analysis \cite{rbofn}, it is  argued that 
0.04 -- 0.2~dex of the observed A(Li) could be post-primordial  in origin.  
On the basis of the very small residual dispersion after accounting  for 
the trend in A(Li) with [Fe/H], and with some ``outliers" removed, RNB 
argue that their data (which may be statistics limited) is consistent with
no dispersion and for an upper limit on the lithium depletion of
0.1~dex.   As  discussed in
\cite{pin1}, the fraction of ``outliers'' is crucial for  constraining
rotationally mixed models.  As of this writing most, if  not all,
evidence points to a rather limited depletion of no more than  0.2~dex,
either in standard stellar models or in those including rotation.

To err on the side of caution, we adopt a central value for the plateau
abundance of A(Li) = 2.2 and we choose a $\sim 2\sigma$ range of $\pm 
0.1$~dex so that our adopted ``95\%" range is: 2.1 $\leq $~A(Li) $\leq 
2.3$.  If depletion is absent, this range is consistent with the lithium 
``valley''. 
For depletion $\geq 0.2$~dex, the consistent lithium 
abundances bifurcate and move up the ``foothills", although a non-negligible contribution from post-primordial lithium could move the primordial abundance 
back down again.

\section{Confrontation Of BBN Predictions And\\
 Observational Data}
 
In the context of the ``standard" model (three families of light or 
massless, two-component neutrinos), the predictions of BBN depend on 
only one free parameter, the nucleon-to-photon ratio $\eta$.  Recalling
that for T$_{0} = 2.728$~K, $\eta_{10} = 273\Omega_{\rm B}h^{2}$, the 
baryon inventory of Persic \& Salucci \cite{ps} may be used to set a very 
conservative lower bound, $\eta_{10} \geq 0.25$.  From constraints on 
the total mass density and the Hubble parameter, the extreme upper bound 
on $\eta$ could be nearly three orders of magnitude larger.  Over this 
large range in cosmologically ``interesting" nucleon abundance, the 
predicted abundance of deuterium changes by more than eight orders of 
magnitude, from more than several parts in $10^{3}$ to less than a part 
in $10^{11}$ as can be seen in Figure 4, where the 
BBN predictions are shown over a wide range in $\eta$.  Over 
this same range in nucleon abundance, the lithium abundance varies from 
a minimum around $10^{-10}$ to a maximum some two orders of magnitude 
larger, while the predicted primordial helium mass fraction 
is anchored between 0.2 and 0.3.  Even the $^3$He abundance, which we 
have set aside due to its uncertain Galactic evolution, varies from much 
higher than observed ($\geq 10^{-4}$) to much less than observed ($\approx 
10^{-6}$).  The key test of the standard, hot, big bang cosmology is to ask 
if there exists a unique value (range) of $\eta$ for which the predictions 
of the primordial abundances are consistent with the light element abundances
inferred from the observational data.  Since we have allowed for the
possibility that one of the two current estimates of primordial deuterium
from extragalactic, absorption studies could reflect the  true abundance 
of primordial deuterium, our test must be done in two parts. Monte Carlo
techniques have proven to be a useful tool in the analysis of  the
concordance between the BBN predictions and the observationally determined
abundances of the light elements \cite{hata,skm} \cite{kr}-\cite{sark2}. 
However, since for the purpose of this review we have taken a broad brush
approach to the observational data, we limit ourselves to a simpler, more 
semi-quantitative discussion of this comparison.

\begin{figure}[h]
	\centering
	\epsfysize=4.5truein 
\hskip .5in 
\epsfbox{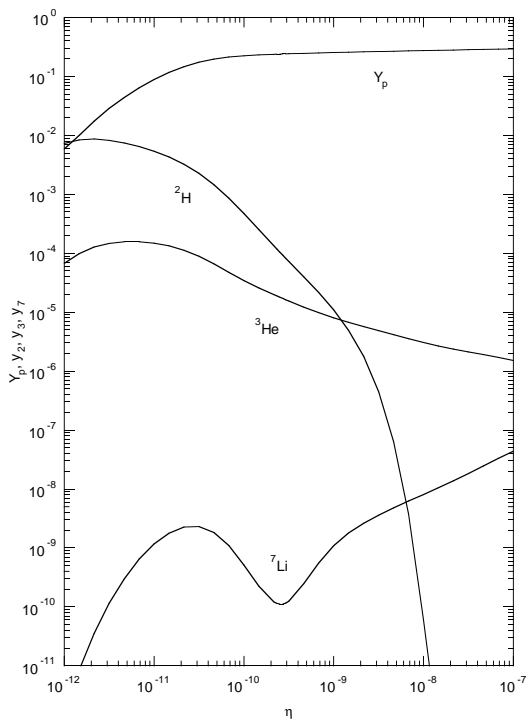}
	\caption{The predicted abundances as a function of $\eta$.}
	\label{fig4}
\end{figure}

\subsection{Low Deuterium}

{}From the two well-observed high-redshift absorption-line systems, we have
adopted the Burles and Tytler 95\% estimate for the primordial-D 
abundance: (D/H)$_{\rm P} = 2.9 - 4.0 \times 10^{-5}$.  With allowance for 
the ($2\sigma$) uncertainties in the BBN-predicted abundance (see Fig. 2), 
the consistent range of $\eta$ is quite narrow: $\eta_{10} = 4.2 - 6.3$.  
For this range in nucleon-to-photon ratio, the primordial lithium abundance 
is predicted (with account for the $2\sigma$ uncertainties in the prediction) 
to lie in the range: A(Li)$_{\rm BBN} = 2.1 - 2.8$.  In our discussion of
the status of  the lithium observational data we identified a range for
its primordial  abundance which has significant overlap with this
predicted range (see  Fig. 3): A(Li)$_{\rm P} = 2.1 - 2.3$.  Thus, the
D-constrained range of 
$\eta_{10} = 4.2 - 6.3$, is consistent with the inferred primordial 
abundance of lithium, even allowing for $\sim 0.2$~dex stellar destruction 
and/or galactic production.  So far, so good.  What of primordial 
helium?  Over this limited range in $\eta$, the predicted helium mass 
fraction varies but little.  With account for the (small) uncertainty in 
the prediction (dominated for this range in $\eta$ by the uncertainty in 
the neutron lifetime): Y$_{\rm BBN} = 0.244 - 0.250$.  This range in the 
predicted primordial helium mass fraction, although on the high side, has 
significant overlap with the range inferred from observations of the
low-metallicity, extragalactic \hii regions: Y$_{\rm P} = 0.228 - 0.248$.  
For ``low-D", the standard model passes this key cosmological test.  
For $\eta_{10}$ in the narrow range from 4.2 to 6.3, the predicted 
and observed abundances of  deuterium, helium-4 and lithium-7 are in 
agreement (and, the predicted abundance of helium-3 is  consistent 
with the abundances inferred  for the interstellar medium and in the 
presolar nebula).

\subsection{High Deuterium}

If, instead, the high abundance of deuterium derived from HST and Keck
observations of one relatively low-redshift absorption-line system is
truly representative of the primordial deuterium abundance, a different
range for the nucleon-to-photon ratio is identified: $\eta_{10} = 1.2 - 
2.8$ (see Fig. 2). The predicted primordial abundance of lithium for 
this range is A(Li) = 1.9 -- 2.7 revealing virtually perfect agreement
with the abundance  derived from the very metal-poor halo stars in the
Spite plateau.  Over  this same $\eta$ range, the predicted helium mass
fraction varies from  Y$_{\rm BBN} = 0.225$ to 0.241.  Here, too, the
prediction is in excellent  agreement with the observed abundance range. 
Thus, for ``high-D" as well,  the standard model passes this key
cosmological test.

\subsection{Consistency With Non-BBN Estimates?}

Having established the internal consistency of primordial nucleosynthesis
in the standard model, it is necessary to proceed to the next key test.
Does the nucleon abundance inferred from processes which occurred during
the first thousand seconds of the evolution of the Universe agree with
estimates/bounds to the nucleon density in the present Universe?

It is a daunting task to attempt to inventory the baryons in the Universe.
Since many (most?) baryons may be ``dark", such approaches can best set
{\it lower} bounds to the present ratio of baryons-to-photons.  In their 
inventory of visible baryons, Persic \& Salucci \cite{ps} estimate for the 
baryon density parameter: $\Omega_{\rm B} \approx 0.0022 + 0.0015h_{50}^
{-1.3}$, where $h_{50}$ is the Hubble parameter in units of 50~km/sec/Mpc.  
For a lower bound of H$_{0} \geq 50$~km/sec/Mpc, this corresponds to a 
lower bound on $\eta$ of: $\eta_{10} \geq 0.25$, entirely consistent with 
our BBN estimates.  More recently, Fukugita, Hogan \& Peebles \cite{fhp} have 
revisited this question.  With subjective, but conservative estimates of 
the uncertainties, their lower bound to the global budget of baryons (for 
H$_{0} \geq 50$~km/sec/Mpc) corresponds to a much higher lower bound: 
$\eta_{10} \geq 1.5$, which is still consistent with the ``low-$\eta$'' 
range we identified using the high D results.  A possible challenge to 
the ``low-$\eta$'' case comes from the analysis of Steigman, Hata \& Felten 
\cite{shf} who used observational constraints on the Hubble parameter, 
the age of the Universe, the ``shape" parameter, and the X-ray cluster 
gas fraction to provide non-BBN constraints on the present density of 
baryons, finding that $\eta_{10} \geq 5$ may be favored over $\eta_{10} 
\leq 2$.  Even so, a significant low-$\eta$, high-D range still survives.

\section{Constraints from BBN}  

Limits on physics beyond the standard model are mostly sensitive to
the bounds imposed on the \4he abundance.  As described earlier, the 
\4he abundance is predominantly determined by the neutron-to-proton 
ratio just prior to nucleosynthesis; this latter is set by the
competition between the weak interaction rates and the universal
expansion rate.  Modulo the occasional free neutron decay, the 
neutron-to-proton ratio ``freezes-out" at a temperature $\sim 800$~
keV.  While the weak interaction rates converting neutrons to protons
and vice-versa are ``fixed", there may be room for uncertainty in the
expansion rate which depends on the total mass-energy density.  For
example, the presence of additional neutrino flavors (or of any other
particles which would contribute significantly to the total energy 
density) at the time of nucleosynthesis would increase the total energy 
density of the Universe, thus increasing the expansion rate, leading 
to an earlier freeze-out, when the temperature and the $n/p$ ratio are 
higher.  With more neutrons available, more \4he can be synthesized.  
In the standard model the energy density at a temperature of order 1~MeV 
is dominated by the contributions from photons, electron-positron pairs
and three flavors of light neutrinos.  We may compare the total energy 
density that in photons alone through N which counts the equivalent
number of relativistic degrees of freedom.
\beq
\rho = (N/2)\rho_{\gamma}
\label{N}
\eeq
In the standard model at $T \sim 1$ MeV, $N_{\rm SM} = 43/4$, so
that we may account for additional degrees of freedom by comparing
their contribution to $\rho$ to that of an additional light neutrino
species
\beq
N = N_{\rm SM} + 7/8 \Delta N_{\nu}
\label {Nnu}
\eeq
For $\Delta N_{\nu}$ sufficiently small, the predicted primordial
helium abundance scales nearly linearly with $\Delta N_{\nu}$:
$\Delta$Y $\approx 0.013\Delta N_{\nu}$.  Hence, any constraints
on Y lead directly to bounds on $\Delta N_{\nu}$ \cite{ssg}.  However, 
it is worth recalling that the constraint is, ultimately, on the ratio
of the Hubble parameter (expansion rate) and the weak interaction rate 
at BBN, so that changes in the weak and/or gravitational coupling constants 
can be similarly constrained \cite{yssr,lim}. Here we will restrict our 
attention to the limits on $N_\nu$ and on neutrino masses from BBN.   
Although likelihood methods have been used to obtain more exact limits 
on $N_\nu$ \cite{nnu}, again here we adopt a simpler, more broad brush 
approach.  Many of the limits on particle properties were recently 
reviewed in \cite{sark1}.

Given the observational upper bound on \Yp of 0.248 and a predicted
lower bound of 0.244 (for low-D), there is room for an increase in the
BBN-predicted \4he of $\Delta$Y = 0.004. From the scaling of Y with 
$\Delta N_{\nu}$, we derive an upper limit to $\Delta N_{\nu}$ of
$\Delta N_{\nu}  < 0.3$. It should be cautioned that this bound is 
really less stringent than a true ``2$\sigma$" upper limit, since we 
have chosen 2$\sigma$ ranges both in the predicted and the observed 
deuterium and helium abundances.  Even so, for low-D this constraint 
is already good enough to permit an exclusion of any ``new", light 
scalars (which would count as $\Delta N_{\nu} = 0.57$), as well as a 
fourth neutrino.  For high-D we predict a lower bound of Y = 0.225 to 
be compared with the observed upper bound of with Y = 0.248, and using 
the same argument, we derive an upper bound of $\Delta N_{\nu} < 1.8$.  

It should be noted that the limit derived above is not restricted to full 
strength weak interaction neutrinos.  In fact, since we know that there
are only three standard neutrinos, the limit is most usefully applied to
additional particle degrees of freedom which do not couple to the Z$^{0}$. 
For very weakly interacting particles which decouple very early, the reduced 
abundance of these particles at the time of nucleosynthesis must be taken 
into account\cite{oss}.  For a new particle, $\chi$, which decoupled at 
$T_d > 1$~MeV, conservation of entropy relate the temperature of the 
$\chi$s to the photon/neutrino temperature (T) at 1~MeV, $((T_\chi/T)^3 
= ((43 / 4N(T_d)))$.  Given $g_{B(F)}$ boson (fermion) degrees of freedom, 
\beq 
\Delta N_{\nu} = {8 \over 7} \sum {g_B \over 2} ({T_B \over T})^4
 +  \sum {g_F \over 2} ({T_F \over T})^4.
\eeq
As an example of the strength of this bound, models with right-handed 
interactions, and three right-handed neutrinos, can be severely constrained 
since the right-handed states must have decoupled early enough to ensure 
that $3(T_{\nu_R}/T_{\nu_L})^{4} \linebreak < \Delta N_{\nu}$.  Using the
high D limit  to $N_\nu$, three right-handed neutrinos requires $N(T_d)
\ga 15$, implying  that $T_d > 40$ MeV.  In contrast, the low D limit
requires that 
$N(T_d) \ga 60$ so that $T_d > 300$ MeV.  If right-handed neutrino interactions 
are mediated by additional gauge interactions, associated with some scale
$M_{Z'}$, and if the right handed cross sections scale as $M_{Z'}^{-4}$, 
then the decoupling temperature of the right handed interactions is related 
to $M_{Z'}$ by $ ({{T_d}_R / {T_d}_L})^3 \propto ({M_{Z'} / M_Z})^4$ which, 
for ${T_d}_L \sim 3$ MeV requires ${T_d}_R \ga 40 (300)$ MeV,  the
associated mass scale becomes 
$M_{Z'} \ga 0.6 (2.8)$ TeV!  Note that this constraint is very sensitive 
to the BBN limit on $N_\nu$. 

Many other constraints on particle properties can be related to the 
limit on $N_\nu$.  For example, neutrinos with MeV masses would also 
change the early expansion rate, and the effect of such a neutrino 
can be related to that of an equivalent number of light neutrinos
\cite{ktcs,kkkssw,dol,fko}.   A toy model which nicely contains ways to 
both increase and decrease \4he production relative to standard BBN 
is the case of a massive \nutau \cite{kkkssw}.  The two relevant 
parameters are the \nutau mass and lifetime.  A \nutau which is stable 
on BBN timescales (\ie, $\tau_\nu \ga 100$ sec) and has a mass greater 
than a few MeV will increase \Yp relative to standard BBN.  This is 
because such a neutrino still decouples when it is semi-relativistic 
and so its number density is comparable to 
that of a massless neutrino.  However, its energy density at the onset 
of BBN is much greater than that of a massless neutrino since its mass 
is significantly greater than the temperature.  Therefore, weak interactions 
decouple earlier, increasing the neutron-to-proton ratio at freeze out 
and thus the amount of \4he.  For example, a limit of $N_\nu \leq 4$ 
translates into a mass limit on a relatively stable (on the time-scale 
of BBN) neutrino of $m_\nu < 0.4$~MeV for a Dirac-mass neutrino and $m_\nu 
< 0.9$~MeV for a Majorana-mass  neutrino \cite{fko}.  Just the opposite 
can occur if such a \nutau decays rapidly  compared to BBN timescales. 
The rapid decays and inverse decays keep the \nutau s in equilibrium 
much longer than do the conventional weak interactions so that their 
number density, along with their energy density, is exponentially suppressed.  
A typical example is a relative decrease in \Yp of about 0.01 for a \nutau 
with a mass of $\sim 10$ MeV and a lifetime (\nutau $\rightarrow \nu_\mu +\phi$ 
where $\phi$ is a Majoron) of 0.1 sec.

\section{Conclusions}

In this precision era of Cosmology the BBN abundances are
predicted with great accuracy in the standard model.  The
statistical uncertainties in the primordial abundances of
the light nuclides inferred from the observational data are
also very small.  However, there is evidence that the derived
abundances may be subject to systematic errors much larger 
than the statistical errors.  This is particularly evident
for deuterium where the D/H ratio derived for two, low 
metallicity, high redshift absorption systems differs by a 
factor of 5 - 10 from that inferred for a third such system.
For \4he, two determinations of the primordial mass fraction
differ from each other by 2 - 3 times the statistical error.
Their differences may be traced to differing treatments of
the corrections for collisional excitation and ionization
and the data sets may be contaminated by some cases of
underlying stellar absorption.  Although a clear, accurately
determined ``plateau" is evident in the Li vs. Fe relation 
for the metal-poor halo stars, the level of the plateau is
subject to uncertainties in the metal-poor star temperature 
scale and atmosphere models.  In addition, there may be
non-negligible corrections (larger than the statistical
uncertainties) due to depletion of surface lithium in these 
very old stars, as well as enhancement due to post-BBN
production.  Nonetheless, despite these nagging uncertainties,
the agreement between the predictions of standard BBN and
the observed abundances is impressive.  The standard model
passes this key test with flying colors.

Given the dichotomy in the possible primordial abundance of
deuterium, we have considered two possibilities.  For the
``low-D" option, we identify a ``high-$\eta$" range (at 95\% 
confidence): $\eta_{10} \approx 4.2 - 6.3$.  In this range
the predicted abundances of \3he, \4he and \7li are consistent
with the primordial abundances inferred from observations (see
Figures 1 -- 3).  For $\eta$ in this range the baryon density
parameter is restricted to: $\Omega_{\rm B}h^{2} \approx 
0.015 - 0.023$ which, for H$_{0} = 70$~km/s/Mpc corresponds 
to: $\Omega_{\rm B} \approx 0.03 - 0.05$.  Using the upper 
bound to \Yp from the data along with the lower bound to 
$\eta$ leads to a ``high-$\eta$" bound to the number of 
``equivalent" light neutrinos: N$_{\nu} \leq 3.3$.  For 
the ``high-D" option a ``low-$\eta$ range is identified: 
$\eta_{10} \approx 1.2 - 2.8$.  In this range as well there 
is overlap between the predicted and observed primordial 
abundances of \3he, \4he and \7li.  For this ``low-$\eta$" 
range, $\Omega_{\rm B}h^{2} \approx 0.004 - 0.010$ which, 
for H$_{0} = 70$~km/s/Mpc corresponds to: $\Omega_{\rm B} 
\approx 0.01 - 0.02$.  In this range the upper bound to the 
number of equivalent light neutrinos is much less restrictive:
N$_{\nu} \leq 4.8$.  As a key probe of early Universe Cosmology
and of particle physics (standard model as well as beyond the
standard model), BBN is alive and well.

\vskip .2in

\noindent {\bf Acknowledgments}

Dave Schramm's impact on the field of BBN and on the lives of the
authors is reflected in and between almost every line of this review.  
The astroparticle connection championed by Dave depends crucially on 
BBN and we are delighted to report in this, his Memorial Volume, that 
BBN remains the fundamental interface between particle physics and 
cosmology.  We can think of few things that would have made him happier.
In addition to Dave, we would like to thank our many collaborators on 
BBN over the years: J. Audouze, T. Beers, S. Bludman, B.A. Campbell, 
M. Casse, C. Chiappini, C. Copi, D.S.P. Dearborn, J.E. Felten, B.D. 
Fields, R. Gruenwald, J. Gunn, N. Hata, C.J. Hogan, K. Kainulainen, 
H.-S. Kang, L. Kawano, M. Kawasaki, P.J. Kernan, E.W. Kolb, H. Kurki-Suonio, 
P. Langacker, G.J. Mathews, F. Matteucci, R.A. Matzner, B.S. Meyer, 
V.K. Naranyanan, J. Norris, M.J. Perry, M.H. Pinsonneault, R. Ramaty, 
R. Rood, S. Ryan,  R.J. Scherrer, S. Scully, E. Skillman, F.K. Thielemann,  
D. Thomas, B.M. Tinsley, M. Tosi, J.W. Truran, M.S. Turner, E. Vangioni-Flam, 
S.M. Viegas, J. Yang.
The work of K.O. was supported in part by DOE grant
DE--FG02--94ER--40823 at the University of Minnesota.
The work of G.S. and T.W. was supported in part by DOE grant
DE--AC02--76ER--01545 at the Ohio State University.

\end{document}